\documentclass[5p,sort&compress,authoryear]{elsarticle}
\usepackage{amsmath}
\usepackage{graphicx,psfrag,epsf}
\usepackage{enumerate}

\usepackage{makecell}
\usepackage[table,dvipsnames]{xcolor}
\usepackage{amsthm}
\usepackage{bm}
\usepackage{bbm}
\usepackage{mathtools}
\usepackage{collcell}
\usepackage{dsfont}
\usepackage{rotating}
\usepackage{wrapfig}
\usepackage[utf8]{inputenc}
\usepackage[T1]{fontenc}
\usepackage{tikz}
\usepackage{lettrine}
\usepackage{dblfloatfix}
\usepackage[english]{babel} 
\usepackage{amsfonts}
\graphicspath{ {fig/} }
\usepackage[hyperfootnotes=false]{hyperref}
\usepackage{subfigure}
\usepackage{lipsum}
\usetikzlibrary{arrows,decorations.markings}
\usepackage{standalone}

\usepackage{tikz}
\usepackage{multirow}

 \definecolor{lgrey}{rgb}{0.8,0.8,0.8}
 \definecolor{lred}{rgb}{1,0.5,0.5}
 \definecolor{lgreen}{rgb}{.3,.7,0.3}
 \definecolor{lblue}{rgb}{.4,0.4,.9}
 \definecolor{lpurple}{rgb}{.7,0.3,.9}
 \definecolor{lorange}{rgb}{.9,0.6,0.3}
 \definecolor{lemph}{rgb}{0.2,0.2,0.6}
 \definecolor{code_comment}{rgb}{0,0.2,0.4}
 \definecolor{code_string}{rgb}{0.4,0.2,0}

\definecolor{gray}{rgb}{0.5,0.5,0.5}
\definecolor{red}{rgb}{0.8,0,0}
\definecolor{dred}{rgb}{0.5,0,0}
\definecolor{blue}{rgb}{0,0.1,1}
\definecolor{dblue}{rgb}{0,0.1,0.6}
\definecolor{cyan}{rgb}{0,0.5,.5}
\definecolor{dcyan}{rgb}{0,0.3,.3}

\definecolor{b}{rgb}{0,0,.8}	
\definecolor{g}{rgb}{0,.6,0}	
\definecolor{n}{rgb}{0,0,0}	
\definecolor{h}{rgb}{0.4,0.2,0.2}	
\definecolor{v}{rgb}{0.2,0.6,0}

\usepackage[normalem]{ulem}
\newcommand{\del}[1]{\color{gray}\ifmmode\text{\sout{\ensuremath{#1}}} \else\sout{#1} \fi \color{black}}

\begin{document}
	\begin{frontmatter}
		\title{Distributional neural networks for electricity price forecasting}
		\author[wroclaw]{Grzegorz Marcjasz}
		\author[essen]{Micha{\l} Narajewski}
		\author[wroclaw]{Rafa{\l} Weron}
		\author[essen]{Florian Ziel}
		\address[wroclaw]{Department of Operations Research and Business Intelligence, Wrocław University of Science and Technology, 50-370 Wrocław, Poland}
		\address[essen]{House of Energy Markets and Finance, University of Duisburg-Essen, 45141 Essen, Germany}
		\begin{abstract}
 			We present a novel approach to probabilistic electricity price forecasting which utilizes distributional neural networks. The 
            model structure is based on a 
            deep neural network that contains a so-called probability layer. The network's
            output is a parametric distribution with 2 (normal) or 4 (Johnson's SU) parameters.
            In a forecasting study involving day-ahead electricity prices in the German market, our approach significantly outperforms state-of-the-art benchmarks, including LASSO-estimated regressions and deep neural networks combined with Quantile Regression Averaging. The obtained results not only emphasize the importance of higher moments when modeling volatile electricity prices, but also -- given that probabilistic forecasting is the essence of risk management -- provide important implications for managing portfolios in the power sector.
  
		\end{abstract}
	\end{frontmatter}
	
	\section{Introduction and motivation}
	
	Trading in competitive markets requires precise probabilistic forecasts. Therefore, the attention of researchers and practitioners is shifting recently from point to probabilistic forecasting methods. It is not different in electricity markets since the liberalization starting in 1990s. The point electricity price forecasting (EPF) literature is very broad, and the topic is well-researched \citep{weron2014electricity}. However, these models can be used in probabilistic forecasting only to a limited extent, as they mostly predict only the expected price, rarely quantiles or other characteristics. A proper risk optimization that is compulsory in the very volatile and uncertain electricity markets can only be carried out using methods that provide a broader view, such as e.g. distributional forecasting. It did not go unnoticed to researchers \citep{nowotarski2018recent, petropoulos2022forecasting}, however the literature on probabilistic EPF is much scarcer than the one on point EPF.
	
	We propose a probabilistic EPF approach based on distributional neural networks. More specifically, we consider a `vanilla' \textit{deep neural network} (DNN), i.e., a multi-layer perceptron in which the information propagates only forward. We utilize the TensorFlow \citep{tensorflow2015-whitepaper} and Keras \citep{chollet2015keras} frameworks, and let the output layer be a parametric distribution with 2 or 4 parameters \citep{salinas2020deepar,barnes2021controlled,barnes2021adding}.
	The difference to a standard network providing point forecasts is only in the output layer. Thus, if we have already built a neural network model for point forecasting, it is very easy to convert it to a distributional one. Even though the method is not new \citep{nix1994estimating,williams1996using}, it has not attracted much attention. To the best of our knowledge, the only existing distributional networks in the energy forecasting literature use mixtures of normal distributions obtained using complex structures comprising convolutional neural networks (CNN) and gated recurrent units (GRU) \citep{afrasiabi2020deep-based} or recurrent neural networks (RNN) \citep{mashlakov2021assessing, brusaferri2020probabilistic}. The \textit{distributional deep neural network} (DDNN) proposed in this paper is far less complex than the CNNs, GRUs and RNNs, easier to interpret and computationally less demanding.

	Neural network-based models are popular and have been shown to achieve good predictive accuracy in the fields of point EPF \citep{keles2016extended,lago2021forecasting}, load forecasting -- both point \citep{yang2018electricity} and quantile \citep{zhou2022electricity} as well as quantile wind power forecasting \citep{he2020probability}. However, despite the relative simplicity of moving from point to probabilistic forecasting using the TensorFlow framework, it has not been utilized in EPF yet. 
 
    The performance of the model is evaluated using a rolling window forecasting study with the day-ahead electricity prices in Germany. The DDNN is benchmarked against naive bootstrapping and two well performing point EPF approaches \citep{lago2021forecasting}: a lasso estimated autoregression (LEAR) and a DNN, both combined with quantile regression averaging (QRA) for converting point predictions into probabilistic ones.
	
	The major contributions of the manuscript are as follows:
	\begin{enumerate}
		\item It is the first work to consider the DDNN and one of the first to consider probabilistic neural networks in electricity price forecasting.
		\item The proposed method is very simple compared to the existing neural network solutions. As we show in this paper, the generalization from a point to a distributional DNN requires almost no effort.
		\item If needed, point forecasts can be easily derived from the predicted distribution. 
		\item It is the first in EPF and one of the first studies to use the Johnson's SU \citep{johnson1949systems} distribution in probabilistic forecasting \citep{mori2021off-block}.
		\item It is a fully automatic forecasting method that may be used with other markets and data. The code is open-source.
		\item The obtained forecasts are interpretable in terms of distribution's characteristics, and the results provide evidence of the importance of higher moments in EPF.
		\item The  aggregation schemes for probabilistic forecasts proposed in the paper yield robust predictions that significantly outperform the state-of-the-art benchmarks in terms of the \textit{continuous ranked probability score} (CRPS).
	\end{enumerate}

    Finally, given that probabilistic forecasting is the essence of risk management, our study provides power market participants with a new, significantly more accurate tool for assessing risks related to trading power portfolios.
 
	The remainder of this manuscript is structured as follows. Section~\ref{sec_prob_ann_descr} introduces the reader to the concept of distributional artificial neural networks. Section~\ref{sec_market} provides an overview of the market and data used in the application study. The models, including the DDNN and the hyperparameter tuning, are described in Section~\ref{sec_models}. The application study together with the results are presented in Section~\ref{sec_application_study}. The paper is concluded with a discussion of the main findings in Section~\ref{sec_conclusion}.

	\subsection{Probabilistic forecasting literature}
	The probabilistic EPF literature is not as rich as the one considering point forecasts, what can be deducted from the reviews of \citet{nowotarski2018recent}, \citet{ziel2018probabilistic} and \citet{hong2020energy}. Researchers consider mainly the day-ahead market, which is the main electricity spot trading place. However in recent years the research focused on other markets as well, e.g. the intraday \citep{uniejewski2019understanding, narajewski2020econometric, narajewski2020ensemble, oksuz2019neural, janke2019forecasting} and balancing markets \citep{kraft2020reserve, browell2022imbalance, janczura2022strategies}. Two widely used and efficient model estimation methods in point EPF are lasso \citep{ziel2016forecasting, ziel2018day, uniejewski2019understanding, narajewski2020econometric} of \citet{tibshirani1996regression} and artificial neural networks \citep{dudek2016multilayer, oksuz2019neural, zhou2019optimized, luo2019two, zahid2019electricity, lago2021forecasting}. A substantial stream of new EPF research considers hybrid models \citep{jahangir2019novel, zhang2020hybrid, oreshkin2021nbeats, olivares2021neural}, however, as \citet{lago2021forecasting} conclude, they often avoid proper comparisons to well-established methods. The probabilistic EPF comprises mostly of quantile regression \citep{maciejowska2016probabilistic, marcjasz2020probabilistic, maciejowska2020assessing}, bootstrapping \citep{efron1979bootstrap} of point forecast residuals \citep{wan2013hybrid, ziel2018probabilistic, narajewski2021optimal}, and of RNNs \citep{mashlakov2021assessing, brusaferri2020probabilistic}. We follow the recommendations of \citet{lago2021forecasting} and compare our model against the very competitive lasso estimated autoregressive (LEAR) and a point deep neural network. These are point forecasting models, thus we apply quantile regression averaging on them. Other methods of obtaining probabilistic forecasts, such as bootstrap, could be considered as well, but we refrain from that for the sake of simplicity. A vast amount of literature concerns also forecast combinations in the electricity markets \citep{hubicka2018note, serafin2019averaging, karabiber2019electricity}, however due to the complexity of the probabilistic forecast aggregation as pointed out by \citet{berrisch2021crps}, we limit ourselves only to two simple averaging schemes with equal weights that allow to stabilize the neural network predictions.
	
	\section{The distributional deep neural network (DDNN) model}\label{sec_prob_ann_descr}
	We assume that the reader is familiar with and understands the concept of the (feed-forward) deep neural networks (DNN). In this section, we briefly recall the definition and the mathematics behind it to underline the difference between the DNN with point and probabilistic output layers. 
	\subsection{Architecture}
	Let $\bm{X} \in \mathbb{R}^{D\times N}$ be the input matrix with $N$ denoting the number of features and $D$ the number of observations. Further, let $\bm{H}_i\in \mathbb{R}^{D\times h_i}$ be the output matrix of $i$th hidden layer, $\bm{W}_i \in \mathbb{R}^{h_{i-1}\times h_i}$ and $\bm{b}_i\in \mathbb{R}^{D\times h_1}$ be the corresponding hidden-layer weights and bias where $h_i \in \mathbb{N}$ is the number of neurons in $i$th hidden layer with $h_0 = N$ and thus $\bm{H}_0 = \bm{X}$. Additionally, denote $a_i(\cdot)$ the $i$th activation function. Then, for $i \in \{1, \dots, I\}$ we have
	\begin{equation}
		\bm{H}_i = a_i\left(\bm{H}_{i-1}\bm{W}_i + \bm{b}_i\right).
	\end{equation}

	Now, we got to the point where the DNN with point and probabilistic output layers differ. That is to say, in the standard DNN we calculate the output $\bm{O}\in \mathbb{R}^{D\times S}$, where $S$ is the number of modelled features. Formally,
	\begin{equation}
		\bm{O} = \bm{H_I}\bm{W}_{I+1} + \bm{b}_{I+1}
		\label{eq_output_layer}
	\end{equation}
	are the values returned by the network. Such DNN is optimized given the true observation matrix $\bm{Y}\in\mathbb{R}^{D\times S}$ with respect to point losses, e.g. the mean squared error (MSE) or mean absolute error (MAE). In the case of the DDNN, the parameter layer $\bm{\Theta}\in\mathbb{R}^{D\times S\cdot P}$ consists of $P$ distribution parameters for each of the $S$ modelled features. It is however computed in the same manner as in equation~\eqref{eq_output_layer}. The final output is made by creating a $D\times S$-dimensional matrix of the assumed distributions $\bm{F}(\bm{\Theta};x)$. The network is then optimized given the true observation matrix $\bm{Y}\in\mathbb{R}^{D\times S}$ with respect to probabilistic losses, e.g. by maximizing the likelihood for a parametric distribution or by minimizing the continuous ranked probability score (CRPS). 
	
\begin{figure*}[t!]
	\centering
	\begin{subfigure}[Deep neural network (DNN) with a multivariate output layer]
		{
			\includegraphics[width=.8\textwidth]{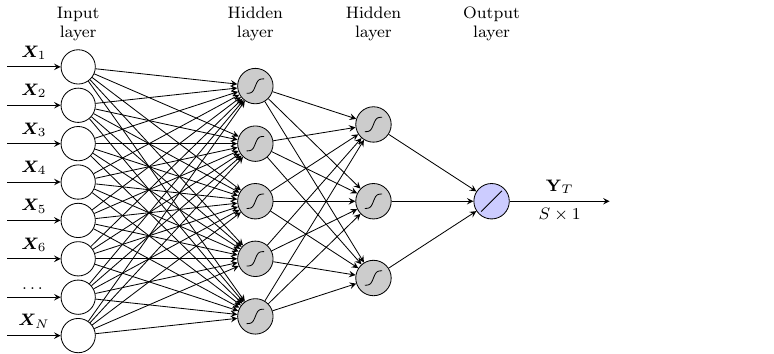}
		}	
	\end{subfigure}
	\begin{subfigure}[Distributional deep neural network (DDNN) with a multivariate output layer]
		{
			\includegraphics[width=.8\textwidth]{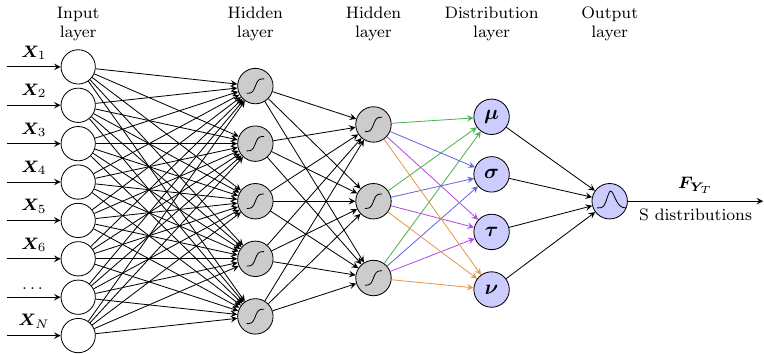}
			\label{fig_ann_bigger_prob}
		}
	\end{subfigure}
	\caption{Comparison of the DNN and DDNN.}
	\label{fig_ann_bigger}
\end{figure*}

	Figure~\ref{fig_ann_bigger} provides an example with $I=2$ hidden layers and this setting we use in the remainder of the manuscript.
The number of neurons in the hidden layers is arbitrary, but the same for both DNNs in order to underline the difference between the point and probabilistic networks. We see clearly that the input and hidden layers are identical for both DNNs and only the output part differs. 

As a  final remark of the subsection, we discuss the multivariate output which consists of multiple features and the possible probabilistic distributions. Namely, we allow in the definition for $S$ output features, and in our setting they are all $S=24$ hours of the electricity prices of the following day. This can be done as all the day-ahead electricity prices are published at once, and therefore they can share the input regressor set. In other applications this is rather not the case, however such a multivariate setting may still be preserved if one considers $S$ similar time series to be forecasted that may benefit from common regressors.

The probabilistic output layer may consist of nearly any implemented probabilistic distribution. Based on application, these can be, e.g., binomial or Poisson if we deal with a discrete problem, gamma or beta if we deal with a continuous problem, but supported only on the positive line, or normal, $t$ or Johnson's SU if supported on the whole real line. As the electricity prices may be both positive and negative, we use in our study the two-parametric normal and four-parametric Johnson's SU distributions. 
	\subsection{Regularization}
	The danger of overfitting the model can be tackled in the DDNN similarly as in the standard one. One could use regularization, a dropout layer or early stopping. We use all of these in our forecasting study, however we approach the regularization of the parameter layer differently.
	
	The DNN design allows for $L_p$ regularization of every hidden layer $\bm{H}_i$, its weights $\bm{W}_i$, and bias $\bm{b}_i$. Applying it to the DNN we get the following loss with regularization
	\begin{equation}
	\begin{aligned}
		\mathcal{L}_{\text{reg}}(\bm{Y}, \bm{O}) &= \mathcal{L}(\bm{Y}, \bm{O}) + \sum_{i=0}^{I} \lambda_{1,i} ||\bm{H}_i||_p + \\
		&+ \sum_{i=0}^{I} \lambda_{2,i} ||\bm{W}_{i+1}||_p + \sum_{i=0}^{I} \lambda_{3,i} ||\bm{b}_{i+1}||_p
	\end{aligned}
	\label{eq_MLP_regularization}
	\end{equation}
	where $||\cdot||_p$ represents the $L_p$ norm. One can flexibly choose between the types of regularization, use both or none, and choose to regularize only some part of the network, e.g., only $\bm{H}_1$ layer and $\bm{W}_2$ weights. The $\lambda_{j,i}$ parameters are subject to hyperparameter tuning. The regularization of the DDNN may be done in the same way as described in Eq.~\eqref{eq_MLP_regularization}, however, we could also regularize each of the distributional parameters separately as follows
	\begin{equation}
	\begin{aligned}
		&\mathcal{L}_{\text{reg}}\left(\bm{Y}, \bm{F}(\bm{\Theta};x)\right) = \mathcal{L}(\bm{Y}, \bm{F}(\bm{\Theta};x)) + \sum_{i=0}^{I-1} \lambda_{1,i} ||\bm{H}_i||_k +\\
		&+\sum_{i=0}^{I-1} \lambda_{2,i} ||\bm{W}_{i+1}||_k + \sum_{i=0}^{I-1} \lambda_{3,i} ||\bm{b}_{i+1}||_k +\\
		&+ \sum_{p=1}^{P} \left(\lambda_{1,I,p}||\bm{H}_I||_k + \lambda_{2,I,p}||\bm{W}_{I+1}||_k + \lambda_{3,I,p}||\bm{b}_{I+1}||_k\right).
	\end{aligned}
	\label{eq_dMLP_regularization}	
	\end{equation}
	The difference between Eq.~\eqref{eq_dMLP_regularization} and Eq.~\eqref{eq_MLP_regularization} is the regularization of the last layer. Namely, in Eq.~\eqref{eq_MLP_regularization} we regularize the whole output layer using the same $\lambda_{j,I}$ values, whereas in Eq.~\eqref{eq_dMLP_regularization} each parameter $p \in \{1,\dots, P\}$ is regularized using its own $\lambda_{j,I,p}$ values. 
	The colour arrows in Figure~\ref{fig_ann_bigger_prob} denote separate kernel $\bm{W}_{I+1}$ regularization for each of the distributions' parameters.
	The reason to use such a  differentiation is the possibility to use different amount of inputs' information for each distribution parameter, what was already observed in the literature \citep{narajewski2020ensemble}.

	\section{The data}\label{sec_market}
	The goal in the empirical case study is forecasting day-ahead electricity prices in Germany. This section familiarizes the reader with the utilized data, especially the input features and the forecasting objective. The electricity markets in Europe consist of derivative, spot and balancing parts \citep{Viehmann2017}. The most important is the spot market, particularly the day-ahead auction. It takes place once a day at noon where all $S$ products of the following day are traded in a uniform  price auction~\citep{weron2019electricity}. In the majority of countries $S=24$, however in some cases like the UK it is $S=48$. As all hours of the following day are traded at once, all of them are based on the same set of information. Therefore, in our study we model all the prices using exactly the same input features, what supports the multivariate output of the DDNN presented in Section~\ref{sec_prob_ann_descr}.
	
	The considered dataset spans six years of hourly observations from 01.01.2015 to 31.12.2020. The study uses a rolling window what mimics the daily business in practice and is a standard procedure in the EPF literature \citep{weron2014electricity, weron2019electricity}. The initial in-sample period spans the date range from 01.01.2015 to 26.12.2018 which consists of $D=4\cdot 364 = 1456$ observations. For the purpose of hyperparameter tuning, we split it additionally to training and validation sets. The out-of-sample period starts on 27.12.2018, and ends on 31.12.2020, however the first 182 observations are used to obtain the QRA forecasts and thus are excluded from the analysis. Therefore, the final out-of-sample test set for probabilistic predictions uses 554 days of data. The models are retrained every day using the most recent $D$ observations and the hyperparameters obtained in the tuning that is run on initial in-sample dataset.
	
\begin{figure*}[p]
		\centering
		\includegraphics[width=\linewidth]{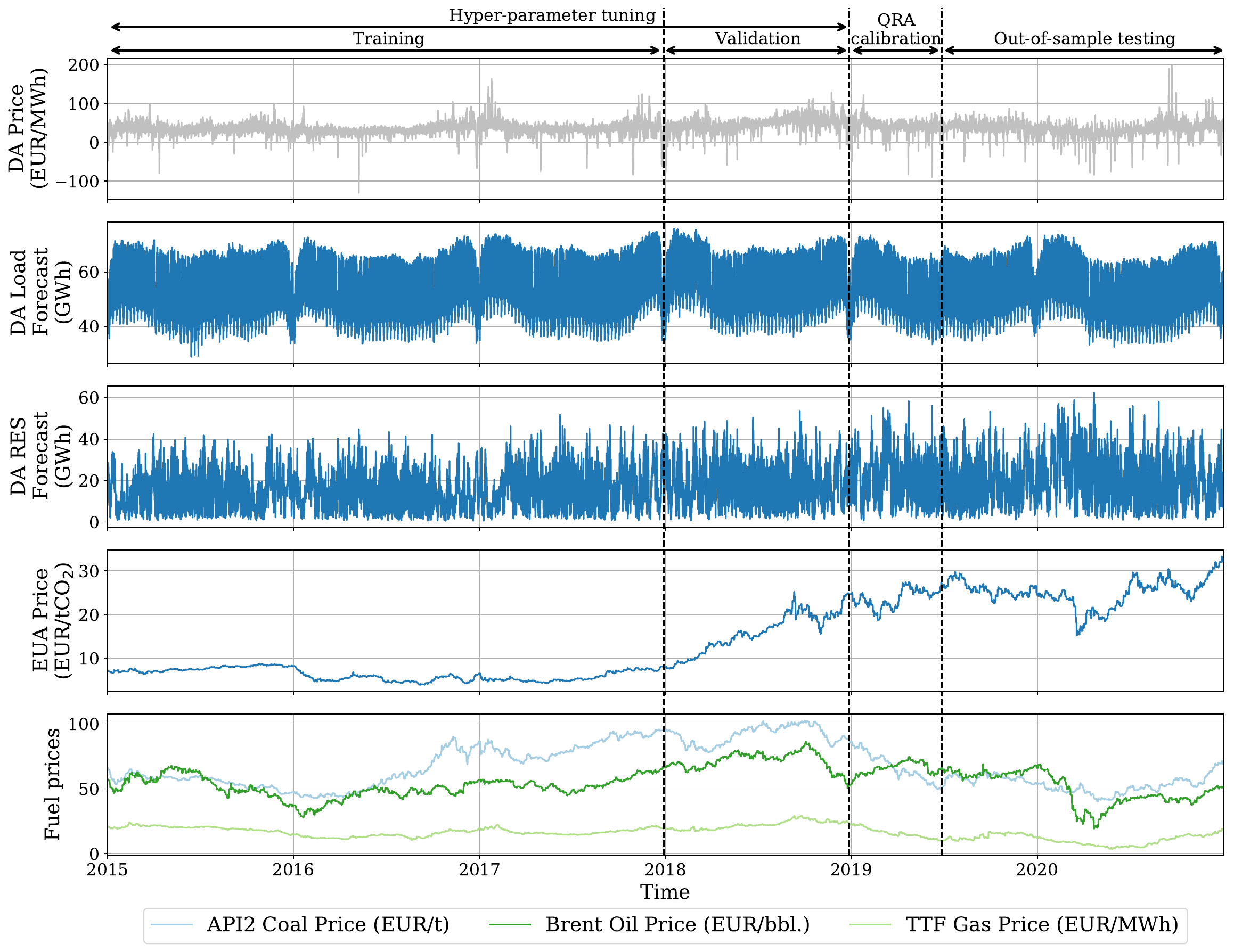}
			\caption{Time series plots of the considered data.}
		\label{fig_data_tsplot}
	\end{figure*}

	Figure~\ref{fig_data_tsplot} shows plots of the considered features together with the dates and study stages mentioned above. 
	The data contains the day-ahead (DA) electricity prices, DA load forecasts, DA renewable energy sources (RES) forecasts, EU emission allowance prices and fuel: coal, oil and natural gas prices. The RES forecast is a sum of wind offshore, wind onshore and solar generation day-ahead forecasts. The DA prices and load forecasts exhibit strong daily, weekly and annual seasonality. Thus,  we model each hour of the day separately within a single neural network and also utilize the weekday dummies. We do not construct any regressor explaining the annual behavior as it is well described by the load data. On the other hand, the RES forecasts exhibit only daily and annual seasonality and the EUA and fuel prices are random-walk type processes. These conclusions might not be easy to derive based on Figure~\ref{fig_data_tsplot}, however see \citet{ziel2018day}, \citet{sgarlato2022role} and \citet{bille2022epf} for more insights.
	
\begin{figure*}[p]
		\centering
		\includegraphics[width=\linewidth]{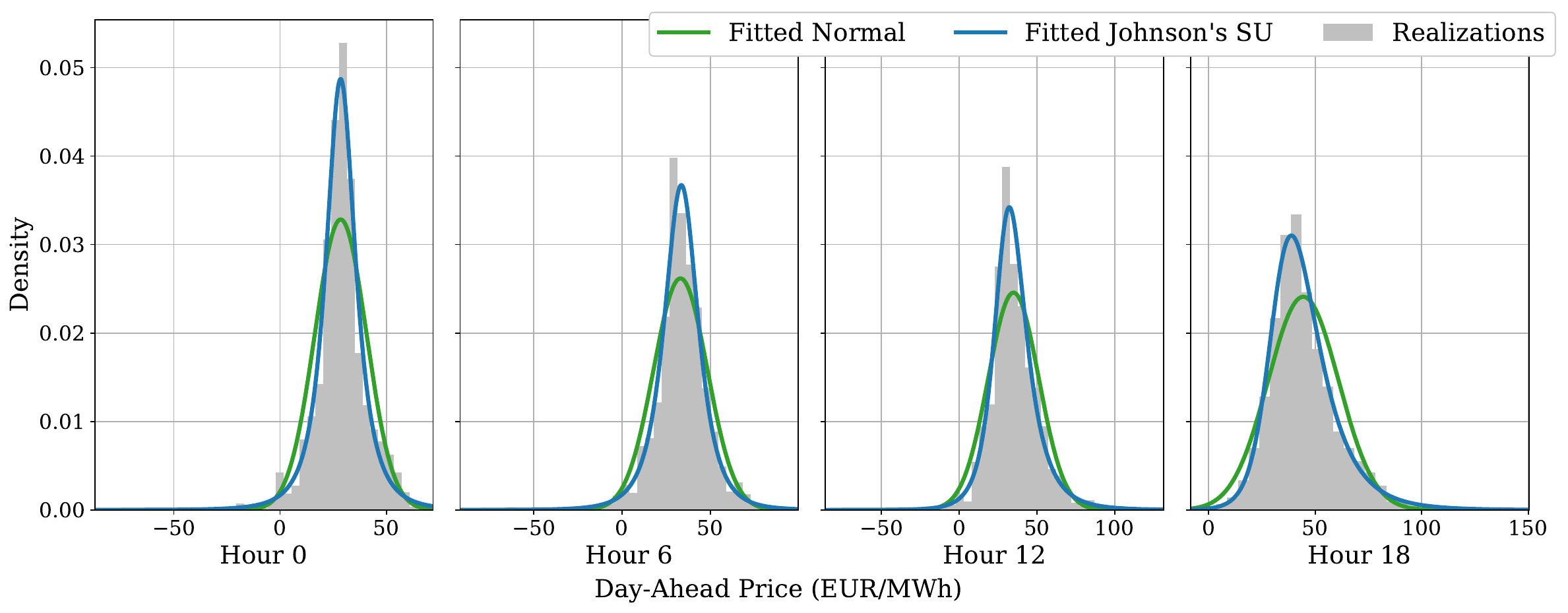}
		\caption{Histograms of prices for selected hours with fitted densities of normal and Johnson's SU distributions. The plots are based on the in-sample (training and validation) data.}
		\label{fig_prices_hist}
	\end{figure*}

	Figure~\ref{fig_prices_hist} presents histograms of prices for selected hours.
	Additionally, we fit there normal and Johnson's SU distributions and plot their densities. Both distributions belong to the location-scale family. The normal distribution $\mathcal{N}(\mu, \sigma^2)$ is a well-known two-parametric distribution with $\mu$ being the location and $\sigma$ the scale parameters. The Johnson's SU distribution $\mathcal{J}(\mu, \sigma, \nu, \tau)$ was first investigated by \citet{johnson1949systems} as a transformation of the normal distribution. It is a four-parametric distribution with $\mu$ being the location, $\sigma$ the scale, $\nu$ the skewness and $\tau$ the tail-weight parameters. 
	So far, it has not found application with distributional neural networks. However, it is often used in the context of energy commodities \citep{patra2021revisiting,gianfreda2018stochastic, abramova2020forecasting}.
	Based on Figure~\ref{fig_prices_hist} we suspect that the Johnson's SU is more suitable for modeling the electricity prices than the normal. We observe heavy tails and skewness what cannot be explained by the normal distribution. Thus, in the forecasting study we use both distributions to emphasize the gain that comes from using the more flexible distribution.

	\section{Models and estimation}\label{sec_models}
	\subsection{Input features}\label{sec_input_features}
	Let us recall that we forecast the $S=24$ day-ahead prices on day $T$, i.e. $\bm{Y}_T = \left(Y_{T,1}, Y_{T,2}, \dots, Y_{T,S}\right)'$. The following input features are available for all considered models:
	\begin{itemize}
		\itemsep0em
		\item Past day-ahead prices of the previous three days and one week ago, i.e. $\bm{Y}_{T-1}$, $\bm{Y}_{T-2}$, $\bm{Y}_{T-3}$, and $\bm{Y}_{T-7}$.
		\item Day-ahead forecasts of the total load for day $T$, i.e. $\bm{X}^L_T = (X^L_{T,1}, X^L_{T,2}, \dots, X^L_{T,S})'$, as well as the past values of the previous day and one week ago, i.e. $\bm{X}^L_{T-1}$, and $\bm{X}^L_{T-7}$.
		\item Day-ahead forecasts of renewable energy sources (RES) for day $T$, i.e. $\bm{X}^{RES}_T = (X^{RES}_{T,1}, X^{RES}_{T,2}, \dots, X^{RES}_{T,S})'$, as well as the past values of the previous day, i.e. $\bm{X}^{RES}_{T-1}$.
		\item EU emission allowance most recent closing price, i.e. $X^{EUA}_{T-2}$. 
		\item Fuels most recent closing prices, i.e. $X^{Coal}_{T-2}$, $X^{Gas}_{T-2}$, and $X^{Oil}_{T-2}$.
		\item Weekday dummies, i.e. $\textbf{DoW}_d(T)$ for $d = 1, 2, \dots, 7$.
	\end{itemize}
	Let us note that the forecasting exercise is performed on day $T-1$ before the day-ahead auction. That is to say, we possess only the information available at around 11:30 CET on day $T-1$. The considered input does not violate this assumption, and therefore we use e.g. $T-2$ lag for the EUA and fuels prices.

	\subsection{Probabilistic neural network}\label{sec_models_nn}
	The probabilistic NN model uses the DDNN described in Section~\ref{sec_prob_ann_descr}. The model consists of 2 hidden layers, $S$ output distributions, and various number of input features. The output distributions are assumed to be either normal or Johnson's SU and each of them defines a separate model. We regularize the model through input feature selection, dropout layer and $L_1$ regularization of the hidden layers and weights. All these are subject to hyperparameter tuning. Additionally, we tune the activation functions, the number of neurons, and the learning rate. The detailed list of hyperparameters and the process is described in Section \ref{sec_hyperparameters}.

	The model is built and estimated using the TensorFlow \citep{tensorflow2015-whitepaper} and Keras \citep{chollet2015keras} frameworks. The hyperparameter optimization is performed with the help of Optuna \citep{akiba2019optuna} package, 4 times for result stability reasons, each time consisting of 2048 iterations. We report the results for each of the 4 optimized hyperparameter sets, as well as for 3 different ensembles of the four distributions, as described in Section \ref{sec_application_study}.
	The model consists naturally also of components that are not tuned in the hyperparameter optimization. That is to say, the model uses additionally an input normalization, negative loglikehood as the loss function, Adam optimizing algorithm, and early stopping callback with patience of 50 epochs. The batch size is fixed to 32, and the maximum number of epochs to 1500. For the rolling prediction, the dataframe was shuffled and $20\%$ was left out for validation.

	Probabilistic neural networks are denoted in the later parts of the paper using \textbf{DDNN-\{distribution\}-\{run\}} scheme, where \textbf{\{distribution\}} is either \textbf{Normal} (or \textbf{N}) or \textbf{JSU} and \textbf{\{run\}} is either a number from 1 to 4 (corresponding to the individual hyperparameter sets), or an indicator of the ensemble of the four: \textbf{pEns} for the vertical average or \textbf{qEns} for the horizontal averaging. See Section \ref{sec_hyperparameters} for the description of different schemes. Note, that when the choice of the distribution is obvious, as in Fig. \ref{fig:pinball}, the \textbf{\{distribution\}} term may be missing in the model acronym.

	\subsubsection{Hyperparameter tuning for neural network models}\label{sec_hyperparameters}

	The neural network models (both point and distributional) underwent the hyperparameter optimization considering below hyperparameters and their potential values:
	\begin{itemize}
		\itemsep0em
		\item Indicator for inclusion of input features described in Section~\ref{sec_input_features} (14 hyperparameters).
		\item Dropout layer -- whether to use the dropout layer after the input layer, and if yes at what rate. The rate parameter is drawn from (0, 1) interval (up to 2 hyperparameters -- the rate is not optimized if dropout layer is not present in the model).
		\item Number of neurons in the hidden layers.  The values are chosen from integers from [16, 1024] interval (1 hyperparameter per layer).
		\item Activation functions used in each of the hidden layers. The possible functions are: elu, relu, sigmoid, softmax, softplus, and tanh (1 hyperparameter per layer).
		\item $L_1$ regularization for hidden layers -- whether to use the $L_1$ regularization on the hidden layers and their weights and if yes at what rate. The rate is drawn from $(10^{-5}, 10)$ interval on a log-scale (up to 2 hyperparameters per layer -- inclusion of $L_1$ for the layer and the rate).
		\item $L_1$ regularization for distribution layer -- separate for each of the $P$ distribution parameters, where $P=2$ for normal and $P=4$ for Johnson's SU distributions -- whether to use the $L_1$ regularization and if yes at what rate (a total of $2P$ hyperparameters; rates identical to the hidden layer regularization).
		\item  Learning rate for the Adam algorithm chosen from the $(10^{-5}, 10^{-1})$ interval on a log-scale (1 hyperparameter).
	\end{itemize}

	The process consists of 2048 iterations of the optimization algorithm which are performed in a hybrid batch-rolling approach. Having the first four years (1456 days) at disposal, we split them into training data (the first 1092 days) and validation data (the last 364 days). Note, that the first day of the out-of-sample test window is the day after the end of the hyperparameter validation data, as illustrated on Figure \ref{fig_data_tsplot} (i.e., there is no data contamination). The hybrid approach is needed to balance two opposing factors. On one hand, a batch estimation (using a single estimation on NN weights) would be less computationally demanding (we would only have 1 neural network trained for every considered hyperparameter set), however the results of such an experiment are very volatile. The best hyperparameter set chosen using the accuracy metric of only a single run would not -- in general -- guarantee a good predictive performance. On the other hand, a rolling setting identical to the one used later (i.e., with a daily recalibration) would be infeasible to compute (as it would take roughly 364 times longer than for batch approach -- we would have 364 neural networks trained for every hyperparameter set). The hybrid approach we have chosen uses 13 recalibrations of neural network models with batches of 28 days estimated using each of the nets. Training data is rolled by 28 days after each step.

	As mentioned earlier, to counteract the local behavior of the hyperparameter optimizer, we repeat the process four times for each of the neural networks. We observe that the predictive performance across the separate hyperparameter sets is not consistent, however the simple aggregation schemes described below provide results consistently better than any of the inputs.

	The first of the aggregation schemes is a mixture distribution, which corresponds to averaging the distributions vertically. However, having two distributions with disjoint pdfs (e.g., two copies of the same distributions significantly shifted), the resulting mixture will be very wide, and might have a ``gap'' in the middle. A more robust alternative is considered,  which utilizes horizontal (quantile) averaging -- i.e., a quantile of an ensemble is computed as an arithmetic mean of the same quantiles from all distributions considered. Such an aggregation in an edge case described earlier would result in an unimodal ensemble distribution, which is much sharper than the vertically averaged one.

	\subsection{Benchmarks}
	\subsubsection{The naive model}
	The first and the simplest benchmark model that we consider is the well-known and widely utilized \citep{weron2014electricity, ziel2018day} \textbf{naive} model. It requires no parameter tuning. Its formula is as follows
	\begin{equation}
		\mathbb{E}\left(\bm{Y}_{T}\right) = \begin{cases}
			\bm{Y}_{T-7},  & \textbf{DoW}_{d}(T) = 1 \,\, \text{for} \,\, d = 1,6,7,\\
			\bm{Y}_{T-1}, & \text{otherwise}.
		\end{cases}
	\end{equation}
	In other words, the \textbf{naive} model uses the prices of yesterday to forecast the prices on Tuesday, Wednesday, Thursday and Friday, and 
	the last week's prices on Monday, Saturday and Sunday. The price distributions are obtained using the bootstrap method which was first proposed by \citet{efron1979bootstrap}. We receive the distributions by adding the in-sample bootstrapped errors to the forecasted expected price
	\begin{equation}
		\widehat{\bm{Y}}_{T}^{m} = \widehat{\mathbb{E}\left(\bm{Y}_{T}\right)} + \widehat{\bm{\varepsilon}}_{T}^m \, \, \text{for} \, \, m = 1, \dots, M
	\end{equation}
	where $\widehat{\bm{\varepsilon}}_{T}^m$ are drawn with replacement in-sample residuals for day $T$, i.e., we sample from the set of $\widehat{\bm{\varepsilon}}_{d} = \bm{Y}_{d} - \widehat{\bm{Y}}_{d}$ for $d = 1, \dots, D$.

    \subsubsection{The LEAR model combined with QRA}\label{sec_LEAR}
	The first of the models that use the structure presented in Section \ref{sec_input_features} is LEAR point forecasting model that uses Quantile Regression Averaging (QRA) to generate probabilistic forecasts. The LEAR model utilizes the LASSO regularization \citep{tibshirani1996regression}. Such an approach eliminates the need for an additional input selection, as the algorithm itself indirectly chooses the most relevant inputs. The regularization parameter (the sole hyperparameter of the LEAR model) is obtained using 7-fold cross validation and a grid of 100 values automatically chosen by a least angle regression (LARS) based estimator. The LEAR approach encompasses a forecast averaging scheme proposed by \citet{lago2021forecasting} -- four independent forecasts are generated for each hour (based on 56, 84, 1092 and 1456 day rolling calibration windows) and the final output is their simple average. Such an approach allows for a balance of the ability to adapt to rapidly-changing market conditions (thanks to the shorter calibration windows) with robustness coming from the use of long windows. It was shown to provide forecasts that -- on average -- are on par or better than all of the comprising forecasts considered separately \citep{lago2021forecasting}.

	There are two ways of using a set of four separate forecasts or an ensemble: one that uses the whole information directly (i.e., the separate forecasts), which we will denote \textbf{QRA} (Quantile Regression Averaging) and \textbf{QRM} (Quantile Regression committee Machine) that uses the ensemble of the point predictions \citep{marcjasz2020combine}.

	Aside from the input data, the QRA and QRM approaches are identical -- both use quantile regression with 182 day rolling calibration window to produce the forecast for each of the 99 percentiles, which approximate the predictive distribution relatively well.

	The LEAR models' results are denoted by \textbf{LEAR-\{CAL\}} for the point forecast estimated using \textbf{\{CAL\}} calibration days (e.g., \textbf{LEAR-1456} for the longest calibration window), \textbf{LEAR-Ens} for an hour-by-hour average of all 4 point forecasts and \textbf{LEAR-QRA} and \textbf{-QRM} for the probabilistic forecasts.

	\subsubsection{The DNN model combined with QRA}\label{sec_pointNN}
	The second set of benchmarks utilizes a point neural network model. It differs from the probabilistic counterpart only in the output construction in the network and hyperparameters corresponding to the missing distribution layer (see Sections \ref{sec_prob_ann_descr}, \ref{sec_models_nn} and Figure \ref{fig_ann_bigger}). The rest of the model setting remains unchanged: DNN model has the same inputs, the same hyperparameter selection and uses the same calibration window lengths and training and validation splits. The loss function for the network is MAE, whereas DDNN uses log-likelihood.
	
	Similarly to the DDNN, for the (point) DNN we also derive four independently-trained hyperparameter sets. This allows us to \emph{i)} measure the robustness of the predictions and \emph{ii)} apply two quantile-regression based methods (QRA and QRM), similarly as for the LEAR point predictions, also using a 182 day rolling calibration window.

	The results are marked with \textbf{DNN-n} for the point forecasts (where $n$ $=1,\,\ldots,\,4$ or \textbf{Ens}) and \textbf{DNN-QRA} and \textbf{DNN-QRM}, respectively for percentile forecasts obtained using quantile regression on the four separate point forecasts and their ensemble.

	\section{Empirical results}\label{sec_application_study}
	Many earlier works show that forecast averaging is often key to achieving accurate predictions. Here, we also aggregate multiple forecast runs to improve the result accuracy and robustness. However, considering probabilistic forecasts instead of the point ones significantly increases the complexity of the aggregation schemes that can be applied. As the detailed discussion is out of scope of this paper, we opted to include only the simple aggregations, based on the equally-weighted averaging or distribution mixing.


	On the probabilistic forecasts side, we have four hyperparameter sets chosen in four separate hyperparameter optimization runs for both the normal and JSU DDNNs. We report the errors of each of them separately, as well as the result of two aggregation schemes: an equally-weighted mixture of the four resulting distributions (vertical aggregation) or a mean of values for a given quantile (horizontal aggregation).


	\subsection{Evaluation}
	While the paper focuses  on probabilistic forecasting, we are also interested in the accuracy of the point forecasts. The latter can be easily derived from the probabilistic ones. Following the best practices of \citet{weron2019electricity}, we report two point-oriented metrics: the mean absolute error(MAE; we use median statistic from the probabilistic methodologies for this metric) and root mean squared error (RMSE; we use mean statistic).

	When it comes to the probabilistic forecasts, we use the CRPS, or rather its approximation -- an average pinball score across 99 percentiles.\citep{gneiting2011quantiles,hong2016probabilistic}:
	\begin{equation}
		\begin{aligned}
		\text{Pinball}&(\hat{Q}_{Y_t}(q), Y_t, q) = \\
		&=
		\begin{cases}
			(1 - q)\left(\hat{Q}_{Y_t}(q) - Y_t\right) &\text{for} Y_t < \hat{Q}_{Y_t}(q),\\
			q\left(Y_t - \hat{Q}_{Y_t}(q)\right) &\text{for} Y_t \geq \hat{Q}_{Y_t}(q),
		\end{cases}
		\end{aligned}
	\end{equation}
	where $\hat{Q}_{Y_t}(q)$ is the forecast of the $q$-th quantile of the price $Y_t$. 

	Additionally for each hour of the day, we perform the Kupiec test \citep{kupiec1995} for unconditional coverage for 50\% and 90\% prediction intervals (PIs). For the CRPS, we aggregate the score across all forecasted hours, whereas for the Kupiec test, we provide the number of hours which passed the Kupiec test.

	Lastly, we use the Diebold-Mariano (DM) test that measures the statistical significance of the difference between the accuracy of the forecasts of two models (here, we use  $A$ and $B$ to discern between them) \citep{uniejewski2019understanding, muniain2020probabilistic, sgarlato2022role}. Let us denote the vector of errors (here, the CRPS scores) of model $Z$ for day $d$ as $L_Z^{d}$. Then, the multivariate loss differential series 
	\begin{equation}
		\Delta_{A,B}^d = ||L_A^d||_1 - ||L_B^d||_1
	\end{equation}
	defines the difference of the $L_1$ norm of loss vectors. For each pair of models, we compute the $p$-value of two one-sided DM tests -- one with the null hypothesis $\mathcal{H}_0: \mathbb{E}(\Delta_{A,B}^d) \leq 0$, which corresponds to the outperformance of model $B$ forecasts by those of model $A$ and the second with the reverse null hypothesis $\mathcal{H}_0: \mathbb{E}(\Delta_{A,B}^d) \geq 0$, complementary to the first one. We use the CRPS as the loss function.
	
	\subsection{Results}
		\begin{figure*}[p]
			\centering
			\includegraphics[width=.9\textwidth,trim={0cm, 0cm, 0cm, 0cm},clip]{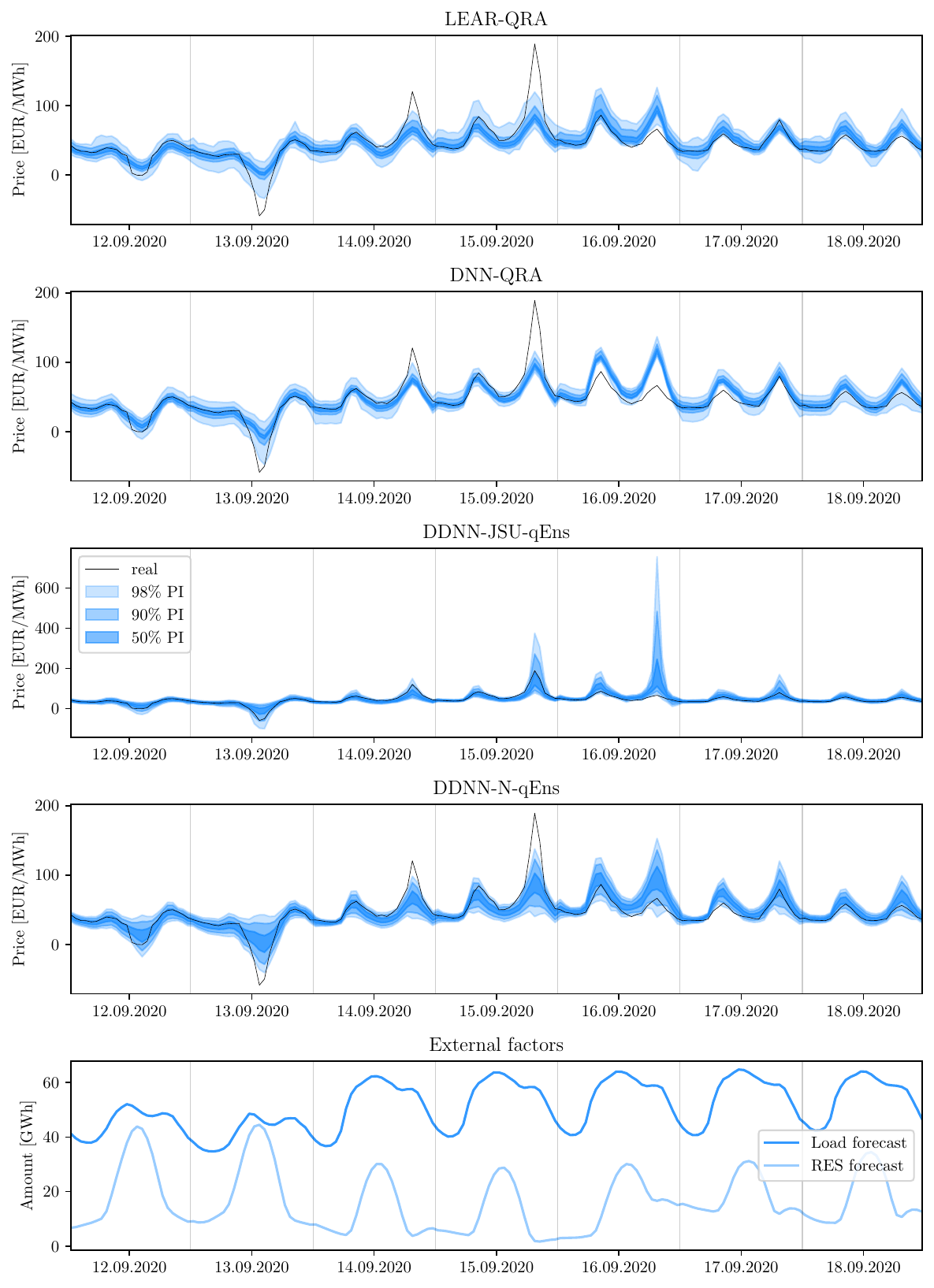}
			\caption{\label{fig:week}Visualization of prediction intervals for a week in September 2020. Quantile (horizontal) averaging with mean is depicted.}
		\end{figure*}

		\begin{figure}[t]
			\centering
			\includegraphics[width=.48\textwidth,trim={.1cm, .35cm, .1cm, .35cm},clip]{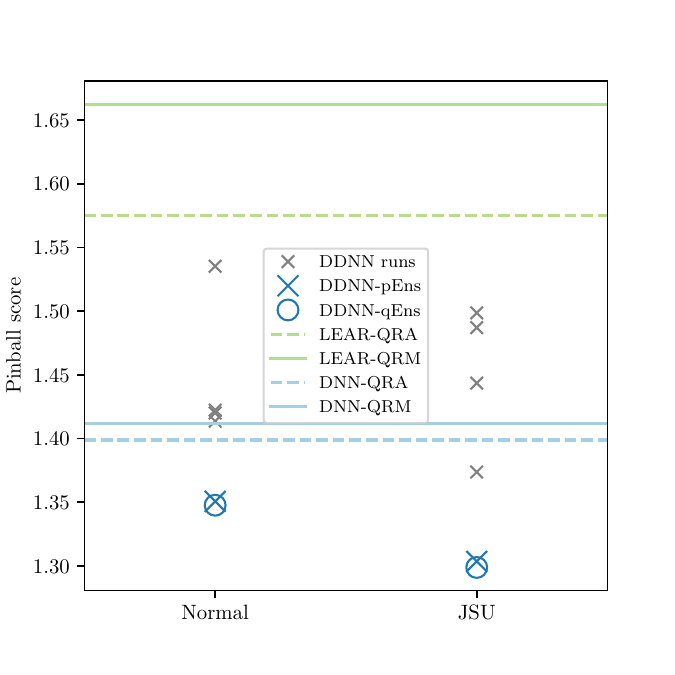}
			\caption{\label{fig:pinball}CRPS scores for benchmarks and DDNNs. Gray markers correspond to the single hyperparameter set results, whereas color ones -- to the combination of the four. Dashed lines mark the QRA method, while solid ones -- QRM.}
		\end{figure}

			\begin{table*}[t]
			  \caption{\label{tab:results}Comparison of point (MAE, RMSE) and probabilistic (pinball, Kupiec test) forecasting accuracy for the considered models.}
			\centering
			\begin{tabular}{rrrrrr}
			  \hline
			 & MAE & RMSE & CRPS & Kupiec 50\% & Kupiec 90\% \\ 
			  \hline
			LEAR-Ens & \cellcolor[rgb]{1,0.5,0.55} {4.372} & \cellcolor[rgb]{1,0.728,0.5} { 6.375} & \cellcolor[rgb]{1,0.5,0.55} {-} & \cellcolor[rgb]{1,0.5,0.55} { -} & \cellcolor[rgb]{1,0.5,0.55} { -} \\ 
			  DNN-Ens & \cellcolor[rgb]{0.813,1,0.5} {3.610} & \cellcolor[rgb]{0.617,0.939,0.5} { 5.850} & \cellcolor[rgb]{1,0.5,0.55} {-} & \cellcolor[rgb]{1,0.5,0.55} { -} & \cellcolor[rgb]{1,0.5,0.55} { -} \\ 
			  naive & \cellcolor[rgb]{1,0.5,0.55} {9.336} & \cellcolor[rgb]{1,0.5,0.55} {14.358} & \cellcolor[rgb]{1,0.5,0.55} {3.585} & \cellcolor[rgb]{0.925,1,0.5} {\textbf{21}} & \cellcolor[rgb]{0.662,0.954,0.5} {\textbf{23}} \\ 
			  LEAR-QRA & \cellcolor[rgb]{1,0.616,0.5} {4.161} & \cellcolor[rgb]{1,0.525,0.5} { 6.676} & \cellcolor[rgb]{0.915,1,0.5} {1.575} & \cellcolor[rgb]{1,0.721,0.5} {10} & \cellcolor[rgb]{1,0.667,0.5} { 8} \\ 
			  LEAR-QRM & \cellcolor[rgb]{1,0.518,0.5} {4.285} & \cellcolor[rgb]{1,0.5,0.55} { 6.788} & \cellcolor[rgb]{1,0.997,0.5} {1.662} & \cellcolor[rgb]{1,0.613,0.5} { 6} & \cellcolor[rgb]{1,0.531,0.5} { 3} \\ 
			  DNN-QRA & \cellcolor[rgb]{0.994,1,0.5} {3.668} & \cellcolor[rgb]{0.597,0.932,0.5} { 5.845} & \cellcolor[rgb]{0.671,0.957,0.5} {1.399} & \cellcolor[rgb]{1,0.613,0.5} { 6} & \cellcolor[rgb]{1,0.721,0.5} {10} \\ 
			  DNN-QRM & \cellcolor[rgb]{1,1,0.5} {3.670} & \cellcolor[rgb]{0.5,0.9,0.5} {\textbf{5.821}} & \cellcolor[rgb]{0.693,0.964,0.5} {1.412} & \cellcolor[rgb]{1,0.694,0.5} { 9} & \cellcolor[rgb]{1,0.667,0.5} { 8} \\ 
			  DDNN-N-pEns & \cellcolor[rgb]{0.978,1,0.5} {3.663} & \cellcolor[rgb]{0.979,1,0.5} { 5.962} & \cellcolor[rgb]{0.588,0.929,0.5} {1.351} & \cellcolor[rgb]{1,0.504,0.5} { 2} & \cellcolor[rgb]{1,0.613,0.5} { 6} \\ 
			  DDNN-N-qEns & \cellcolor[rgb]{1,1,0.5} {3.670} & \cellcolor[rgb]{0.978,1,0.5} { 5.962} & \cellcolor[rgb]{0.583,0.928,0.5} {1.348} & \cellcolor[rgb]{1,0.802,0.5} {13} & \cellcolor[rgb]{1,0.992,0.5} {20} \\ 
			  DDNN-JSU-pEns & \cellcolor[rgb]{0.5,0.9,0.5} {\textbf{3.542}} & \cellcolor[rgb]{1,0.882,0.5} { 6.146} & \cellcolor[rgb]{0.508,0.903,0.5} {1.304} & \cellcolor[rgb]{1,0.5,0.523} { 1} & \cellcolor[rgb]{1,0.558,0.5} { 4} \\ 
			  DDNN-JSU-qEns & \cellcolor[rgb]{0.604,0.935,0.5} {3.564} & \cellcolor[rgb]{1,0.863,0.5} { 6.174} & \cellcolor[rgb]{0.5,0.9,0.5} {\textbf{1.299}} & \cellcolor[rgb]{1,0.829,0.5} {14} & \cellcolor[rgb]{1,0.802,0.5} {13} \\ 
			   \hline
			\end{tabular}
			\end{table*}

		In terms of the CRPS scores, as can be seen in Figure \ref{fig:pinball} and Table \ref{tab:results}, the LEAR-based methods are significantly worse than the neural network-based approaches. However, the performance of the latter is not robust -- run-to-run, the CRPS scores differ by as much as 10\%. As discussed in Section \ref{ssec:multihyper}, this is not known \emph{ex-ante}, therefore an aggregation scheme is needed. After ensembling, regardless of the aggregation scheme applied (vertical, horizontal with mean, horizontal with median), we see similar performance. The normally-distributed networks yield a CRPS of ca. 1.35, whereas JSU ones -- 1.30, i.e., ca. 3-4\% better than the normal. DNN-QR methods can be placed between the probabilistic DDNN ensembles and the individual runs.



		As shown in Table \ref{tab:results}, the neural network-based models are better than LEAR-based ones also for the point forecasts. Interestingly, the ensemble of DNN forecasts has the third lowest error -- both in terms of MAE and RMSE. The best model according to MAE is DDNN-JSU-pEns, followed by its -qEns counterpart -- the two models with the best CRPS score. However, these models are worse w.r.t. the RMSE than all other NN-based models. On the other hand, we see the lowest RMSE for DNN-QR based methods, closesly followed by the DNN-Ens model. The DDNNs are ca 2\% (normal) and 7\% (JSU) worse. There are only minor differences between the vertical and horizontal aggregation schemes.

	Additionally, we performed the Kupiec test for unconditional coverage with 5\% significance level for 50\% and 90\% prediction intervals. From what can be seen in Table \ref{tab:results}, QRA seems to perform better than QRM -- for both the LEAR and point DNN models. However, the QR-based approahces pass the Kupiec test for at most 10 hours of the day. The probabilistic DDNNs, on the other hand show mixed performance. The p-Ens forecasts are worse than most other methods, while q-Ens are better than QR-based predictions. As the p-Ens models sport the CRPS scores similar to the q-Ens ones, the latter are a much better choice, especially when chosen with a more robust median quantile instead of mean. Note, however, that the worst overall method (Naive benchmark) provides the best coverage for both 50 and 90\% PIs.

\begin{figure*}[t!]
	\centering
	\includegraphics[width=.8\textwidth]{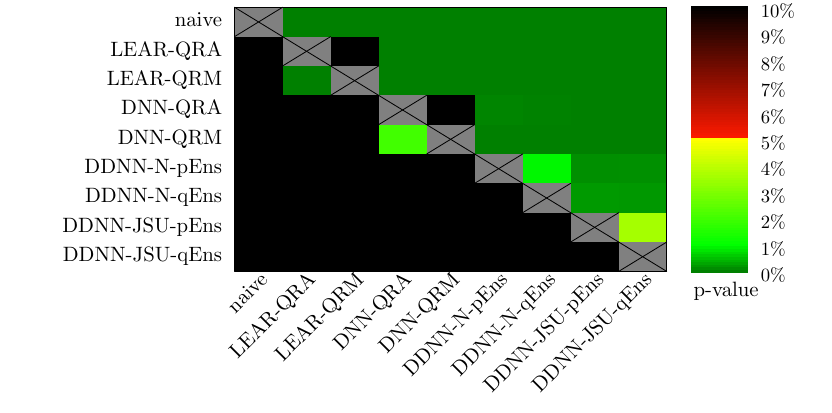}
	\caption{Results of the Diebold-Mariano test. The plots present p-values for the CRPS loss --- the closer they are to zero ($\to$ dark green), the more significant the difference is between forecasts of X-axis model (better) and forecasts of the Y-axis model (worse).}
	\label{fig:dm_test}
\end{figure*}

	Lastly, the results of the DM test are visualized in Figure \ref{fig:dm_test}. We can observe that DDNN-JSU-qEns is the best model overall, with forecasts significantly better than from any other model (represented by a column with all cells green or yellow). Secondly, the DDNN-N-qEns is also significantly better than both other aggregation schemes. Lastly, QRA models (both LEAR and DNN ones) produce significantly better forecasts than their QRM counterparts.

	\subsection{The need for multiple hyperparameter sets}
	\label{ssec:multihyper}

\begin{figure*}[b!]
	\centering
	\includegraphics[width=.8\textwidth]{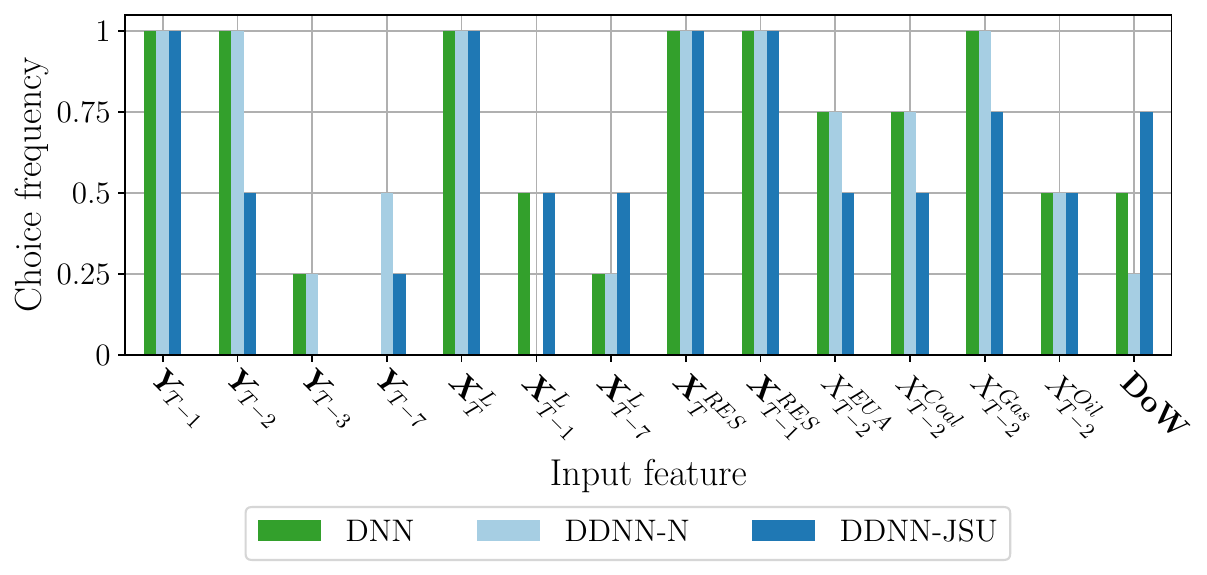}
	\caption{The frequency with which (out of 4 hyperparameter optimization trials) an input group was chosen for inclusion in DNN models. The groups are described in detail in Section \ref{sec_input_features}}
	\label{fig:input_freq}
\end{figure*}

	Even though the hyperparameter optimization uses a repeated neural network training procedure to mimic the rolling calibration window setting used later for the evaluation, the optimal sets obtained using independent hyperparameter trials differ significantly. Moreover, all the optimal sets have a similar, i.e. within 2\% difference, \emph{in-sample} error metric -- what is not reflected in the out-of-sample error obtained using this set. Here, the differences are much more prominent, up to 10\%. The locality of the hyperparameter optimization is clearly visible in the optimal sets chosen in independent trials, despite most of the trials being stalled after around 1000 iterations. Figure~\ref{fig:input_freq} shows choice frequency of the considered input features (i.e., the number of hyperparameter sets that uses a particular input variable), described in Section~\ref{sec_input_features}. All 3 considered neural network models are quite consistent when selecting the inputs. The most important ones are the prices of the previous day and two days ago, the current DA forecasts of load and RES, the previous day's DA forecasts of RES and the recent gas price. The least important are the further lags of prices and load forecast.

	Besides the differences in the inputs chosen, hidden layer sizes are the most prominent, especially for the probabilistic networks. They found optima in both the larger and smaller networks, as shown in Table~\ref{tab:hyperpar_layers}. For example, one of the probabilistic neural networks that used the JSU distribution uses 565 and 962 neurons in the hidden layers (amounting to over 540,000 weights just between the two hidden layers), whereas other had 940 and 58 (over 54,000 weights) or 123 and 668 (over 82,000 weights). Moreover, even the activation functions chosen were not unanimous, but \texttt{softplus} seems to be the best for the first hidden layer. We also observe that the dropout is almost never chosen, similarly to the regularization of the network weights.	
	\begin{table*}[t!]
		\caption{Activation functions and number of neurons selected in each of the hyperparameter tunings.}
			\centering
			\resizebox{\textwidth}{!}{
		\begin{tabular}{|cr|rrrr|rrrr|rrrr|}
		  \hline
		  & & \multicolumn{4}{c|}{DNN} &  \multicolumn{4}{c|}{DDNN-N} &  \multicolumn{4}{c|}{DDNN-JSU}\\
		 & Run &    1 &    2 &    3 &    4 &   1 &   2 &   3 &   4 &      1 & 2 &      3 &      4 \\
		  \hline
		\multirow{2}{*}{Layer 1} & Activation &  softplus &  softplus &  softplus &  softplus &  softplus &  softplus &  softplus &  softplus &  softplus &  elu &  softplus &  softplus \\
		& Neurons    &       906 &       912 &       979 &       965 &       948 &       266 &       542 &       110 &       565 &  940 &       243 &       123 \\
		\hline
		\multirow{2}{*}{Layer 2} & Activation &  softplus &  softplus &      relu &       elu &       elu &      relu &  softplus &      relu &      relu &  elu &       elu &       elu \\
		& Neurons    &       901 &       619 &       767 &       448 &       554 &       883 &       641 &       823 &       962 &   58 &       895 &       668 \\
		  \hline
		\end{tabular}
		}
		\label{tab:hyperpar_layers}
	\end{table*}

	To conclude, we observe that there is a need for repeating the hyperparameter optimization process. Despite the robust optimization setting, the end results are vastly different -- both in terms of the parameters chosen, and the out-of-sample error metrics. A form of the forecast combination is crucial for the outperformance of QR-based methods.

	\section{Conclusions}\label{sec_conclusion}
	
	The paper proposes an application of distributional neural networks to probabilistic day-ahead electricity price forecasting and a simple, yet well-performing aggregation scheme for the distributional neural networks that stabilizes the predictions. Since  probabilistic forecasting is the essence of risk management -- Value-at-Risk (VaR) is nothing else but a quantile forecast -- our study provides important implications for managing portfolios in the power sector.
 
    Comparing the results with the literature approaches, we observe a strong performance of the neural networks -- both the probabilistic forecasts from the proposed methods and from quantile regression applied to their point counterparts are significantly more accurate than the statistical-based combination of LEAR and quantile regression. The added complexity of the neural network having to model the distribution of the data, rather than just their expected values, proves effective, especially when the limitations incurred by the distribution choice are not too severe.
	
	Interestingly, the benefit of using distributional neural networks is visible also when mean absolute errors of the median (50th percentile) forecasts are considered. The DDNN-JSU-Ens approach is the only one that outperforms the ensemble of point NNs in this regard.

	\section*{Acknowledgments}
	This research was partially supported by the Ministry of Science and Higher Education (MNiSW, Poland) through Diamond Grant No. 0219/DIA/2019/48 (to G.M.) and the National Science Center (NCN, Poland) through grant No. 2018/30/A/HS4/00444 (to R.W. and F.Z.).
	
\bibliographystyle{elsarticle-harv}

\small
\bibliography{bibliography}

\end{document}